\documentclass[12pt]{article}
\pdfoutput=1
\usepackage{geometry} 
\usepackage{authblk}
\usepackage{crop}
\usepackage{amsmath,amssymb,amsfonts,upref,endnotes}
\usepackage{amsthm}
\usepackage{graphicx}
\usepackage{color}
\title{The thermodynamic efficiency of computations made in cells across the range of life}
\author[1]{Christopher P. Kempes}
\author[1,2,3]{David Wolpert}
\author[4]{Zachary Cohen}
\author[5]{Juan P{\'e}rez-Mercader}
\affil[1]{The Santa Fe Institute}
\affil[2]{Massachusetts Institute of Technology}
\affil[3]{Arizona State University}
\affil[4]{University of Illinois Urbana-Champagne}
\affil[5]{Harvard University}

\begin{document}

\maketitle

\begin{abstract}
Biological organisms must perform computation as they grow, reproduce, and evolve. Moreover, ever since Landauer's bound was proposed it has been known that all computation has some thermodynamic cost -- and that the same computation can be achieved with greater or smaller thermodynamic cost depending on how it is implemented. Accordingly an important issue concerning the evolution of life is assessing the thermodynamic efficiency of the computations performed by organisms. This issue is interesting both from the perspective of how close life has come to maximally efficient computation (presumably under the pressure of natural selection), and from the practical perspective of what efficiencies we might hope that engineered biological computers might achieve, especially in comparison with current computational systems. Here we show that the computational efficiency of translation, defined as free energy expended per amino acid operation, outperforms the best supercomputers by several orders of magnitude, and is only about an order of magnitude worse than the Landauer bound. However this efficiency depends strongly on the size and architecture of the cell in question. In particular, we show that the {\it useful} efficiency of an amino acid operation, defined as the bulk energy per amino acid polymerization, decreases for increasing bacterial size and converges to the polymerization cost of the ribosome. This cost of the largest bacteria does not change in cells as we progress through the major evolutionary shifts to both single and multicellular eukaryotes. However, the rates of total computation per unit mass are nonmonotonic in bacteria with increasing cell size, and also change across different biological architectures including the shift from unicellular to multicellular eukaryotes.
\end{abstract}

\footnote{This article represents a revision from an earlier version originally submitted on January 10, 2017 to the Philosophical Transactions of the Royal Society A.}

\section{Introduction}
At the center of understanding the evolution of life is identifying the constraints faced by biological systems 
and how those constraints have varied across evolutionary epochs. For example,
a question that often arises in evolutionary theory is how relevant the contingent constraints faced by modern life are 
for understanding 
early life or even the origin of life. Another example is the question of how organisms cope with the constraints
of distinct physical scales --- a dependence that by definition does \emph{not} 
change across evolutionary epochs.

The laws of thermodynamics restrict what biological systems can do on all physical scales and in all evolutionary epochs (e.g. \cite{rasmussen,england,mehta,lynch}). In addition, all known living systems perform computations.
Accordingly, the deep connection between computation and the laws of thermodynamics are a fundamental constraint operating on life across all physical scales and evolutionary epochs \cite{england,mehta,lynch}. 
This implies that by analyzing the thermodynamic properties of biological computation, and in particular the efficiency of those computations,
we may gain insight into the changing constraints that have governed the evolution of life.

A deeper understanding of the thermodynamics of biological systems
may also help to address a question that pervades almost all of biology: how to quantify the fitness of organisms in a more nuanced way than by their instantaneous relative reproduction rates. One way to make progress on this question is to
understand the more fundamental processes that govern reproduction rates. In this regard, it is
worth noting that recent research has
derived reproduction rates (growth rates) of organisms from their metabolic power budgets, thus illustrating the deep connection between energetic efficiency, the cost of organism functions, and reproductive success (e.g. \cite{westogm,kempes}). 
Another way to make progress is to analyze other important organism functionalities in addition
to reproduction rate. Here too thermodynamics is vitally important. For example, 
important organism features such as the tapering of vascular network structure can be predicted from considerations of minimizing energy dissipation \cite{west,west2}. Clearly then, analyzing biological systems in terms of their thermodynamic efficiency can provide insight on how to quantify the fitness of organisms. This approach to defining ``fitness' is analogous to recent efforts that have defined the concept of ``genes'' and ``functionality'' in terms of chemical computations \cite{stadler}. 

In this paper we extend previous work on thermodynamics in biological systems in several ways. We begin by discussing the surprising ways that the overall thermodynamic efficiency of different biological architectures, quantified as power per unit mass, varies across both physical scale and the age of first appearance. These shifts in overall thermodynamic efficiency provide the backdrop against which we compare shifts in the computational thermodynamic efficiency within the cell across biological architectures. First we consider the thermodynamic efficiency of the cellular computation of copying symbolic strings during translation 
in a single ribosome. We then consider the \emph{useful} thermodynamic efficiency of this computation at the whole cell level, 
i.e., the total energy rate for all translation, including protein replacement, divided by the rate of translational done for replication.
This calculation requires consideration of intracellular decay processes and ribosome and protein scaling. After a consideration
of the computation of translation, we consider the efficiency of both the computation of DNA replication and
of maintaining storage capacity, from the scale of cells up to the biosphere. 

\section{Power usage across biological scales}

\begin{figure}[h!]
\begin{minipage}[b]{.7\linewidth}
\text{(a)\quad\quad\quad\quad\quad\quad\quad\quad\quad\quad\quad\quad\quad\quad\quad\quad\quad\quad\quad\quad\quad\quad\quad\quad}
\text{}
\centering\includegraphics[width=1.\textwidth]{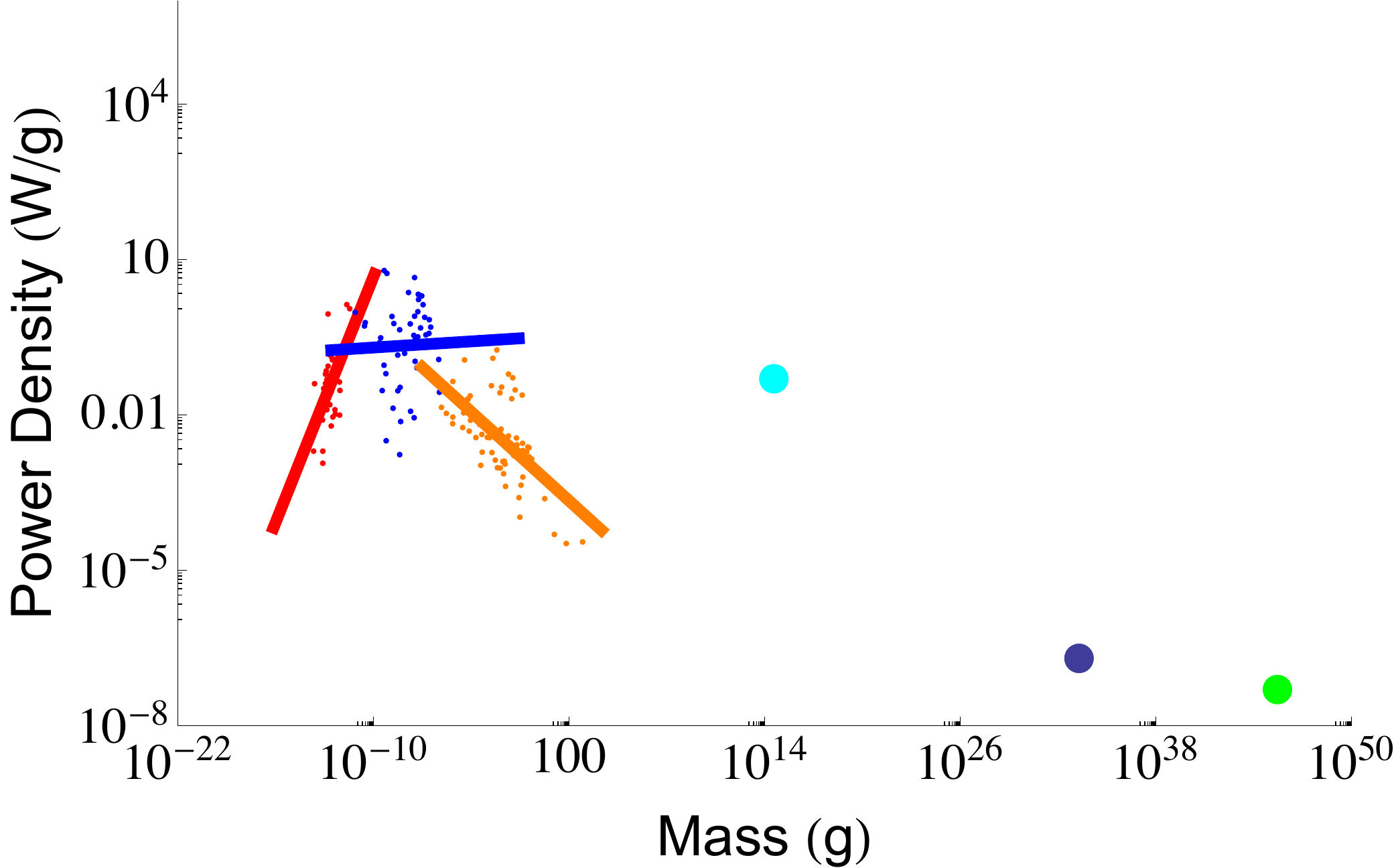}
\end{minipage}
\begin{minipage}[b]{.7\linewidth}
\text{(b)\quad\quad\quad\quad\quad\quad\quad\quad\quad\quad\quad\quad\quad\quad\quad\quad\quad\quad\quad\quad\quad\quad\quad\quad}
\text{}
\centering\includegraphics[width=1.\textwidth]{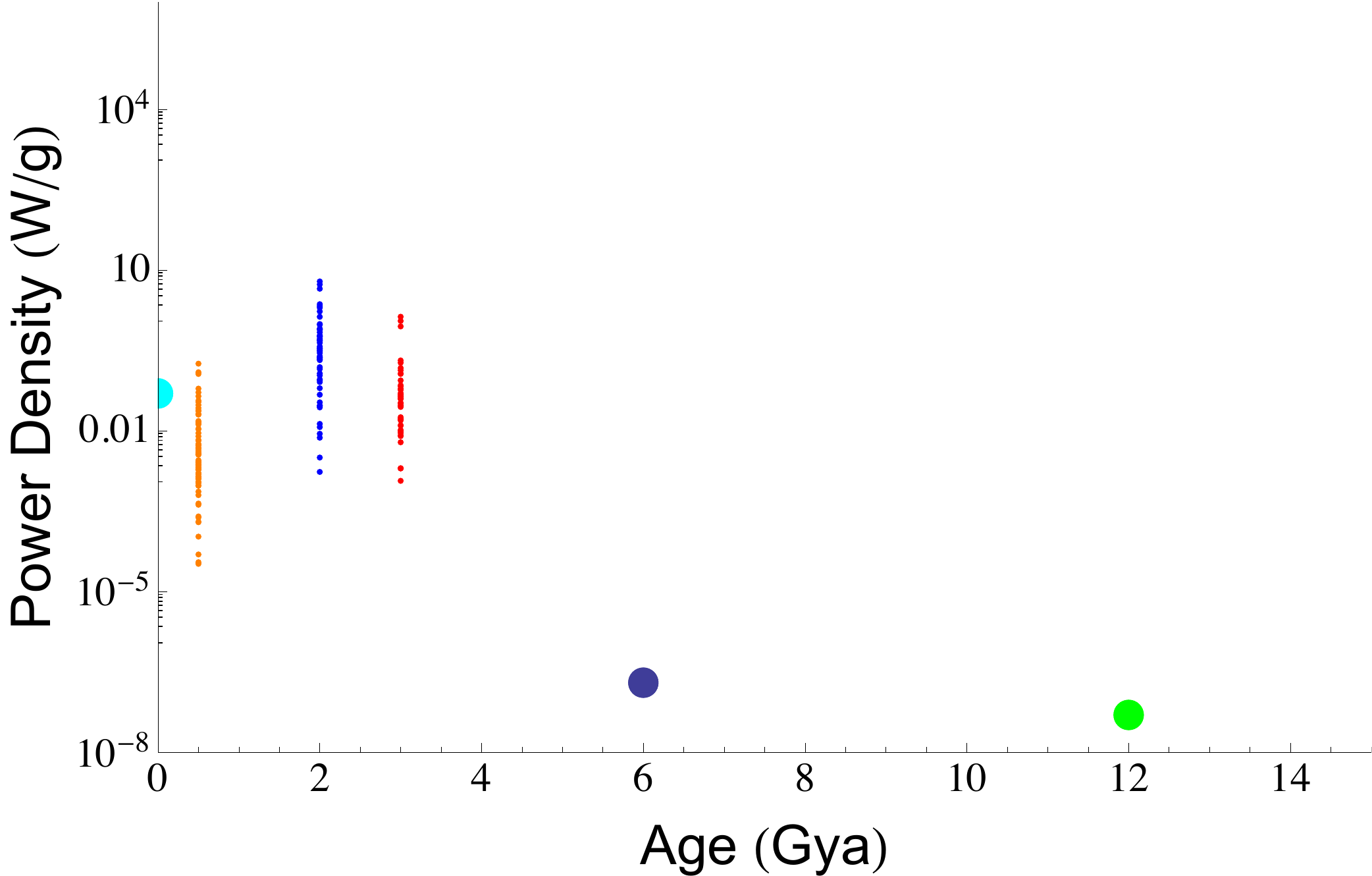}
\end{minipage}
\caption{a.) The overall power density of various systems in the universe. The red points and 
associated red power-law curve are bacteria, the blue points and curve are unicellular eukaryotes, 
and the orange is for multicellular eukaryotes, cyan is modern human society followed by the sun in purple and the milky way in green. b.) Power density as a function of the age (years before the present) of first appearance of each system. The data shown are a reanalysis of \cite{delong,kempes,chaisson,chaisson2}.}
\label{power-density}
\end{figure}

In this paper we frequently investigate features of efficiency by using power laws of the form $Y=Y_{0}X^{\beta}$ where $Y_{0}$ is a normalization constant, $\beta$ is the scaling exponent, and $X$ represents the scale of the system. This type of 
equation makes it easy to relate behaviors at different scales. For example, if $\beta=1$ then all changes in $Y$ are simply proportional to changes in $X$. Similarly, for $\beta \neq 1$ the ratio $Y/X$ will not be constant and will either increase or decrease with increasing $X$.

Before turning to the thermodynamics of biological computation specifically, it is useful to consider the scaling of total organismal power usage in order to gain insight on the bulk thermodynamic efficiency of distinct biological architectures. This perspective will allow us to separate total power usage from the cost of computational rates within cells, and to distinguish which features are changing (or not) across the evolution of life. 

A surprising feature of life at the multicellular scale is that overall metabolic rate does not simply scale linearly with total body size. This is traditionally know as Kleiber's law \cite{kleiber,west}, expressed as a power law with $\beta\approx 3/4$. This
value of $\beta$ implies that multicellular life obeys a certain economy of scale: as organisms grow larger the metabolic rate required to support a unit of mass is decreasing and larger mammals support more tissues for the same amount of energy (e.g. \cite{west}). More recently it has been observed that this scaling relationship is not preserved across all the taxa of life \cite{delong}. In bacteria $\beta$ is greater than $1$ and in unicellular eukaryotes the exponent is close to, but slightly smaller than, $1$ \cite{delong}.  
These relationships imply fundamentally different scaling behavior for each taxa. They have also been used to derive interspecific growth trends and the limits for the smallest possible bacterium and largest unicellular eukaryote \cite{kempes,kempes2}. 

Other work has extended a consideration of power usage from the scales of life to comparisons with astronomical objects.
This work has argued that a characteristic of the aging universe is the appearance of structures with ever higher power density (power per unit mass) \cite{chaisson,chaisson2}. (We will return to this specific claim below.)

The scaling relationships mentioned above can also be used to analyze power density across the epochs of life, since
the noted differences in scaling relationships imply very different power efficiency across each form of life. For example,
bacteria require an increasing amount of power to support a unit of mass with increasing cell size, but are able to reproduce more quickly as a consequence \cite{delong,kempes}. More generally, in the evolution of life power-density first increased with increasing size (bacteria), 
then saturated for unicellular eukaryotes, and then decreases with size for multicellular organisms. It also decreases with size in astronomical systems such as the sun and milky-way (Figure \ref{power-density}a). The surprising features here are: 
\begin{enumerate}
\item the opposing power density relationships for bacteria compared with multicellular life;
\item multicellular organisms fall along a power density curve that would fall below astronomical systems at the same scale implying they would be more efficient (in the sense of requiring less power density to maintain themselves);
\item human societies are well above the average curve for multicellular organisms implying a possible inefficiency -- humanity
is extremely profligate, using power for more than simple maintenance.
\end{enumerate}
In addition, Figure \ref{power-density}b provides the power density as a function of the estimated time when each group of systems arose (in the same way done in \cite{chaisson,chaisson2}) and reveals that the biological groups largely overlap independent of the time of first appearance. Considering the averages of each group, the surprising feature is that the evolution of biological architecture first increased and then decreased the power density as a function of first appearance. 

These observations highlight a critical question: how should we interpret power-density? 
Phrased informally, should a species be proud or ashamed of its power density? The answer ultimately comes down to how effectively power density is converted into functionality. This is a challenging question to address, both because {\it function} is often hard to quantify in terms of increased survival, and because it varies widely across species. In bacteria, we know that the overall power usage predicts the appropriate growth rates from a partitioning between biosynthesis and repair costs \cite{kempes}. However it has been previously noted that this comes at
the cost of a lower efficiency of biomass production \cite{delong} compared with unicellular and multicellular eukaryotes. 
More generally, organisms with a wide variety of growth rates and biomass production efficiencies exist in nature. This either highlights very different selective pressures in different environments (e.g. classic r/K selection theory \cite{macarthur,pianka}), or that there are other quantifications of functionality that are more uniform across diverse species. 

The thermodynamics of computation, which we consider in the rest of this paper, provides a potential starting place 
for analyzing this issue of function.

\section{The Thermodynamics of Computation}

In this section we first provide background on the modern understanding of the thermodynamics
of computation, grounded in nonequilibrium statistical physics. We then discuss the different
kinds of computation that take place in biological systems, clarifying the (very narrow) set of
computations that we consider in this paper.

\subsection{Formalizing and generalizing Landauer's bound} 

There has been great interest for over a century in the relationship between thermodynamics and 
computation~\cite{szilard1964decrease,bril62,wiesner2012information,still2012thermodynamics,prokopenko2013thermodynamic,prokopenko2014transfer,zurek1989thermodynamic,zure89b,bennett1982thermodynamics,lloyd1989use,lloyd2000ultimate,del2011thermodynamic,fredkin1990informational,fredkin2002conservative,toffoli1990invertible,leff2014maxwell}.
A breakthrough was made with the semi-formal arguments of Landauer, Bennett and
co-workers that there is a minimal thermodynamic cost of $kT \ln[2]$
required to run a 2-to-1 map like bit-erasure on any physical system~\cite{landauer1961irreversibility,landauer1996minimal,landauer1996physical,bennett1973logical,bennett1982thermodynamics,bennett1989time,bennett2003notes,maroney2009generalizing,plenio2001physics,shizume1995heat,sagawa2009minimal,dillenschneider2010comment,fredkin2002conservative}.
A related conclusion was that a 1-to-2 map
can act as a \emph{refrigerator} rather than a heater, \emph{removing} heat from the environment~\cite{bennett1989time,bennett2003notes,landauer1961irreversibility,bennett1982thermodynamics}. For example, this occurs in adiabatic demagnetization of an Ising spin system~\cite{landauer1961irreversibility}.

More recently, there has been dramatic progress in our formal understanding of
non-equilibrium statistical physics and its relation to information-processing in general~\cite{faist2012quantitative,touchette2004information,sagawa2012fluctuation,crooks1999entropy,crooks1998nonequilibrium,chejne2013simple, jarzynski1997nonequilibrium,esposito2011second,esposito2010three,parrondo2015thermodynamics,sagawa2009minimal,pollard2014second,seifert2012stochastic,dillenschneider2010comment,takara_generalization_2010,prokopenko2015information,hasegawa2010generalization,takara_generalization_2010}.
In particular, to focus on the specifically computation-based thermodynamic cost of a process,
suppose that at any given time $t$ all states $x$ have the same energy. 
It is now known that in this situation
the minimal work required to transform a distribution $P_0(x)$ at time $0$ to a 
distribution $P_1(x)$ at time $1$ is exactly 
\begin{eqnarray}
kT [S(P_0) - S(P_1)]
\label{eq:1}
\end{eqnarray}
where $S(.)$ is Shannon entropy and $x$ lives in a countable space $X$. 
This lower bound on the work is achieved if and only if the process implementing the
transformation is thermodynamically reversible~\cite{hasegawa2010generalization,parrondo2015thermodynamics,maroney2009generalizing}.
This theoretical result is now being confirmed
experimentally~\cite{dunkel2014thermodynamics,roldan2014universal,berut2012experimental,koski2014experimental,jun2014high}. 
(If the Hamiltonians are not uniform at both times then the change in expected value of the Hamiltonian
must be added Eq.~\eqref{eq:1}.)

This recent work -- and in particular~\eqref{eq:1} -- has fully clarified the early 
reasoning of Landauer et al. To see how,
suppose that the state space $X$ is binary, $P_0(x)$ is uniform, and $P_1(x)$ is a delta function
about $x = 0$. So the transformation is bit erasure, with a uniform initial distribution
of the state of the bit. For this special case, the bound in~\eqref{eq:1} 
giving the minimal work is just $kT \ln[2]$, Landauer's bound. Note that for a different
initial distribution $P_0(x)$, the minimal work will be less than $kT \ln[2]$. More importantly,
note that the bound in~\eqref{eq:1}  
is achieved with a thermodynamically \emph{reversible} process; in general,
logical irreversibility and thermodynamic irreversibility need not imply one 
another~\cite{maroney_absence_2005}. (Indeed, if we use a thermodynamically
irreversible process to implement the logically irreversible map from uniform 
$P_0(x)$ to delta function $P_1(x)$, then the total work used exceeds Landauer's bound of $kT \ln[2]$.)

Viewed as a computation, bit erasure has the very special
property that its output (namely the value 0) is independent of its input. Obviously
this is not true for the vast majority of computations that we might wish to implement in the real world; 
in almost all computations of interest the output depends crucially on the input. 
Moreover, the analyses that have been used to derive~\eqref{eq:1} implicitly
exploited this feature of bit erasure; they only
work for physical  processes in which the output is independent of
the input. This restriction means we
cannot use those analyses to analyze more general types of computation.

To rectify this, in~\cite{dhw_2015_arxiv} a physical process was analyzed that can implement an
arbitrary computation in a thermodynamically reversible way, even a computation
whose output depends on its input. (See also~\cite{wolpert_baez_entropy.2016,maroney2009generalizing}.) That analysis established that~\eqref{eq:1} still
applies for arbitrary computations. Importantly, this result --- which we call the \textbf{generalized Landauer bound} ---
holds even if the conditional distribution $P(x_1 \mid x_0)$ is not a single-valued
map. Indeed, that conditional distribution is not directly relevant; only the resultant marginal distribution
$P(x_1) = \sum_{x_0} P(x_0) P(x_1 \mid x_0)$ arises in the generalized Landauer bound.

\subsection{Thermodynamics of Biological Computation}

\begin{figure}[h!]
\centering\includegraphics[width=.75\textwidth]{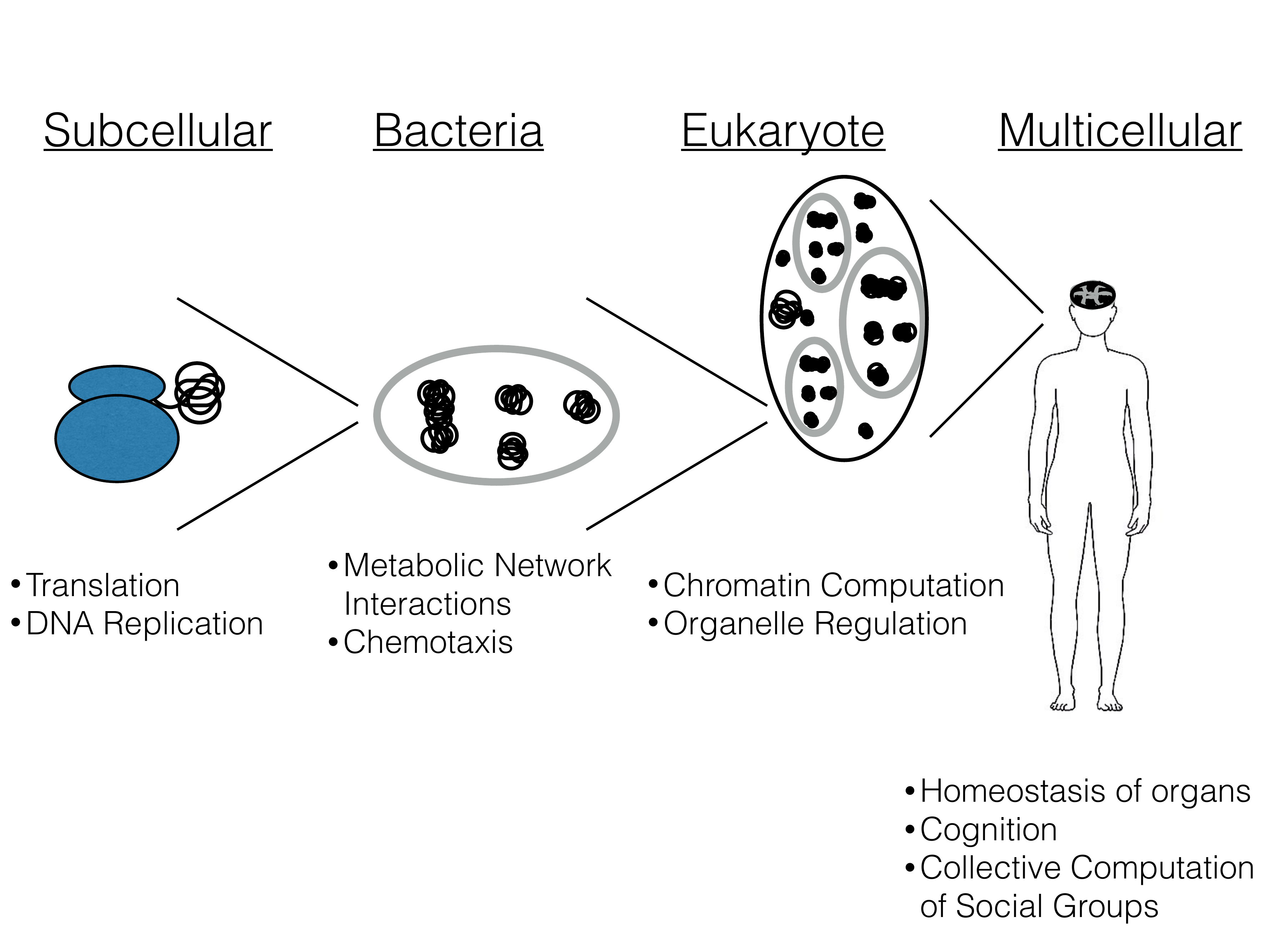}
\caption{The hierarchies of biological computation ranging from subcellular processes to 
interactions within collections of multicellular organisms. At each level of biological organization a few examples of the dominant computational processes are provided. It should be noted that the highest levels of biological computation integrate the lower levels \cite{flack}. For example, while neurons are integrated into a complicated cognitive process they also contain the string writing processes of basic protein translation.}
\label{biohierarchy}
\end{figure}

In artificial digital computers, there is no uncertainty about what precise dynamical process constitutes
a ``computation''. Things are not so clear-cut with biological systems however. One natural criterion
is to characterize a biological process as a ``computation'' if it is reliable, repeatable, and especially
if it can clearly be modeled as a digital operation \cite{shinar,wolpert_baez_entropy.2016}. We adopt this criterion here,
viewing any system that meets these criteria as performing a ``computation''. (At a mininum, we feel
that such systems perform ``computation'' at least as much
as does a binary system undergoing a 2-to-1 map, which ever since Landauer has been
viewed as a canonical model of a computation.)

However even having settled on a clear criterion for what constitutes a biological computation,
there is still a major challenge in accurately identifying and then counting
all the computation operations in a biological system. This is because so many computational processes operate at so many levels of biological organization (see, for example, \cite{flack,flack2} for a review, discussion, and formalism). For example, simply counting the number of
bit operations in the multiple interactions of any given chemical in a metabolic network is
quite difficult, let alone the bit operations involved in controlling all cellular processes such as uptake rates, chemotaxis, and metabolic regulation. Similarly, for human cognition it is not only necessary to consider the input-output operations of 
each separate neuron, but the ``software'' of overall human cognition, which of course involves vast numbers of neurons and is yet to be fully understood.

Adopting a broad perspective, we 
can loosely think of each organism as having their own biological computational hierarchy,
where each level in the hierarchy contains all of the lower levels of computation,
combining them to form unique higher level computations.  For example the bacterial cell contains the basic computations of the ribosome, but also combines the produced proteins to run more advanced computations at the level of metabolic networks. Similarly, a mammal contains computations ranging from translation within its individual eukaryotic cells up to signaling networks coupling those cells, and at a higher level, participates in social organization \cite{flack,flack2}. Thus a full treatment of the energetics of computation requires not only understanding how power usage 
varies  over different levels of biological organization (as we addressed above), but understanding the
associated hierarchy of computations straddling those levels. 
Figure \ref{biohierarchy} illustrates this nesting and gives examples of the new types of computation at each level of biological organization. 

A full treatment of this nested computation is a huge task, which will be central to future work in the emerging field of biological computation \cite{flack,flack2}. Here we only present results based on the most conservative counting of the number of bit operations performed by several of the simplest biological processes, focusing on translation and replication. 
These results are meant to provide intuition on the overall efficiency of the cell, but by no means account for all bit operations in the cell. In addition, our approach only provides a lower bound on the thermodynamic expenditures of those processes that we
do consider, modeling those processes as the writing of one-dimensional strings, without considering the additional
thermodynamic costs of those processes associated with changes in particle number or positional entropies.

\section{Efficiency of cellular computation}

In this section we start by considering the thermodynamic efficiency of translation, first by analyzing the efficiency
of the ribosome as the basic unit for translation, and then by analyzing the efficiency of translation for replication at the level of the
entire cell, which includes the overhead costs of protein replacement. We pay particular attention to how these efficiencies vary with the
size of the cell, and therefore across evolutionary scales. We then broaden our scope to consider other ways of measuring how the amount
of computation varies with size. We conclude by scaling up our analysis to consider the thermodynamic
efficiency of the translational computation performed by the biosphere.

\subsection{Thermodynamic efficiency of computation of translation}
\label{sec:raw_translation}

First, we consider the energetic efficiency of translation, which represents the simple computation of writing free-floating amino acids into distinct strings. Translation is a particularly well-defined biological computation because it produces a repeatable output (the polymerized amino acid chain) from a given input (mRNA) with a high degree of reliability, which as we argued above qualifies it
as a computation. (It also should be noted that this computation is more complex than the simple bit erasure 
considered in analyses of the Landauer bound.) 

We have long known that it takes 4 ATP for the ribosome to add an amino acid to the growing protein chain \cite{lever,lynch,hopfield}. There has been much past theoretical and empirical work on the energetics of translation ranging from arguments of kinetic proof reading \cite{hopfield} to the ratio of forward and backward reaction rates \cite{england}. However, each of these contexts relies on knowledge of the actual chemical process of translation either in terms of natural reaction timescales or the free energy change of certain reactions \cite{hopfield,england}. Our goal here is to compare translation to any physical process performing the same abstract operation. It should be noted though that the expenditure of these 4 ATP is partitioned into two key steps: charging the tRNA, which requires 2 ATP, and forming the peptide bond between amino acids which requires another 2 ATP (e.g. \cite{lever,lynch}). These two chemical processes 
are used together to take the amino acids out of solution and bind them together. 

Biological translation is a computation that writes a specific string of length $l_{p}$ using a $20$ letter alphabet.
It is achieved with a specific chemical process involving tRNA, mRNA, amino acids, and the ribosome. We are interested in quantifying how thermodynamically
efficient this chemical process is by comparing its thermodynamic cost to the smallest possible thermodynamic cost that would
be incurred by {\emph any} physical process that performs the same computation. The generalized Landauer bound provides us with
precisely such a ``scale'' for assessing the  thermodynamic efficiency of biological translation compared with all possible processes performing the same operation.

We can apply Eq.~\eqref{eq:1} to calculate the minimal free energy required to implement the many-to-one mapping 
that transforms a pool of free-floating amino acids (a bath of a large number of uniformly distributed amino acid abundances) into the prescribed amino acid sequence of a 
particular protein. Since there is only one final state, the final entropy is $S_{F}=0$, and if we generalize slightly to a scenario of
$C$ possible amino acids, we get
\begin{equation}
S_{I}=-\sum_{i=1}^{N} p_{i} \ln \left(p_{i}\right).
\end{equation}
In particular, if we have a uniform $p_{i}$, i.e., $p_{i}=p=1/C^{l_{p}}$, then there are a total of $N=C^{l_{p}}$ states, and $S_{I}=Np\ln\left(p\right)=\ln\left(1/C^{l_{p}}\right)$. In many of the calculations below $C$ is taken to be $20$, the actual number of amino acids, we have left the value general in some of our formulas so that considerations of a reduced amino acid pool, although unlikely, could be considered.

Given that the average protein length is $\bar{l}_{p}=325$aa (see \cite{kempes2} for a review of values) for $20$ unique amino acids, we have that $p_{i}=p=1/20^{325}=1.46\times10^{-423}$ where there are $20^{325}$ states such that the initial entropy is $S_{I}=20^{325}p \ln \left(p\right)$ which gives the free energy change of $kT(S_{I}-0)=4.03\times10^{-18}$ (J) or $1.24\times10^{-20}$ (J/ amino acid). This value provides a minimum for synthesizing a typical protein. We can also calculate the biological value from the fact that if $4$ ATP is required to add one amino acid to the polymer chain with a standard free energy of $47.7$ (kJ/mol) for ATP to ADP, then the efficiency is $1.03\times10^{-16}$ (J) or $3.17\times10^{-19}$ (J/ amino acid).  This value is about $26$ times larger than the generalized Landauer bound. 

It should be noted that the efficiency of the translation system is much closer to the Landauer bound than estimates for other biological processes. For example, synapses have been estimated to be $10^{5}$ to $10^{8}$ times worse than the Landauer bound \cite{laughlin-neuron}. 

These results illustrate that translation operates at an astonishingly high efficiency, even though it is still fairly far away from the Landauer bound. To put these results in context, it is interesting to note that the best supercomputers perform a bit operation at roughly $5.27\times10^{-13}$ (J) \cite{landenmark,laughlin}. In other words, the cost of computation in supercomputers is about 7 orders of magnitude worse than the Landauer bound of $kT\ln\left(2\right)=2.87\times10^{-21}$ (J), which is about \emph{5 orders of magnitude less efficient}
than biological translation. Biology is beating our current engineered computational thermodynamic efficiencies by an
astonishing degree. 

There are subtleties in defining the set of states of the system undergoing translation, and therefore the associated changes in entropy. The calculation above of the change in entropy during string writing is based on the specific situation of an infinite bath of uniformly distributed amino acids. We note that although this is a reasonable zeroth-order approximation of the cellular environment, it is an underestimate of the thermodynamic efficiency of translation, since
only some of the entropic costs of translation have been accounted for. In particular, the entropic cost to pull amino acids from
a three-dimensional bath into a one-dimensional string is neglected. Thus, there are other entropy accountings that are worth investigating, including one that begins to approximate the three dimensional problem. Below we provide a few of these alternatives.

First, consider the case where a particular protein is being written from a pool of only the exact amino acids (both number and composition) required for that protein. In this case the number of distinguishable states is given by
\begin{equation}
m=\frac{l_{p}!}{\prod_{k=1}^{C}\left(n_{k}!\right)}
\label{multinom}
\end{equation}
where here $C$ is the number of distinct amino acids used, $n_{k}$ is the number of amino acid of type k, and $\sum_{k=1}^{C}n_{k}=l_{p}$. The initial entropy is then calculated using $S_{I}=mp\ln\left(p\right)$ with $p=1/m$. In the case where all amino acids are used in equal proportion, this alternative would be $m=l_{p}!/\left[\left(l_{p}/20\right)!\right]^{20}$ which, using the values above, gives a Landauer bound of $3.86\times10^{-18}$. This value is very close to the bound calculated above for the case of an infinite uniform bath. The maximal thermodynamic cost in Eq. \ref{multinom} decreases as the number of amino acid types employed decreases. The smallest non-zero value is given by employing two amino acid types with only a single amino acid from one of the types, in which case $m=l_{p}!/\left[\left(l_{p}-1\right)!\right]$. This gives a Landauer bound of $2.40\times10^{-20}$ which is about  two orders of magnitude smaller than the original bound above.  

As we have mentioned, this process of string writing underestimates the change in entropy because it does not account for the three-dimensional positional entropies of each amino acid. One way to start to address this challenge, which still underestimates the positional entropy, is to consider each individual amino acid as distinguishable from the others regardless of type. Under this condition $m=n!$, and the Landauer bound would be $6.46\times10^{-18}$ which is about $1.6$ times the original estimate from the uniform bath. 

These calculations do not prove that evolutionary fitness is highly dependent on thermodynamic efficiency; conceivably,
the majority of chemical processes that perform an analogous computation have such an efficiency. 
Indeed, the specific energetics of translation are tied to the precise processes operating in the ribosome. 
It is at least conceivable that evolution, or an alternate origin of life, could have found an
even more efficient chemical process that operates even closer to the Landauer bound. Nonetheless, these calculations
are at least consistent with the hypothesis that the evolutionary fitness of cells has been highly dependent on the thermodynamic
efficiency with which they perform the computation of translation.

\begin{figure}[h!]
\centering\includegraphics[width=.75\textwidth]{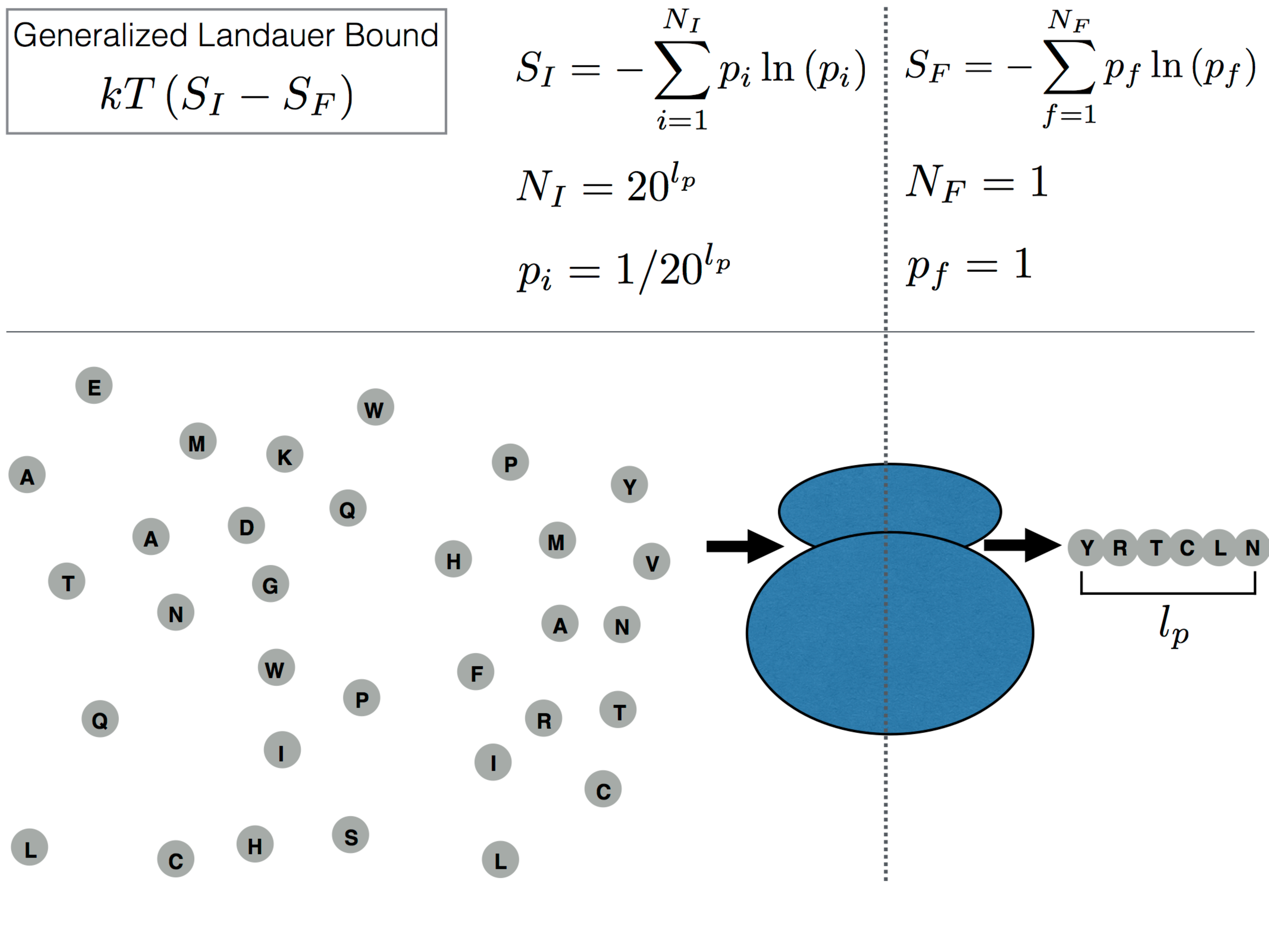}
\caption{Minimal free energy required for the string writing process of protein translation, taking a uniform bath of amino
acids to a specific protein. The formula for the generalized Landauer bound is given, along with the initial and final entropies of writing a specific string of length $l_{p}$ (the average length of a protein in amino acids) from a set of twenty objects (amino acids).}
\label{translation-schematic}
\end{figure}

\subsection{Thermodynamic efficiency of useful translation}

The calculations in Sec. 4~\ref{sec:raw_translation} 
do not involve the size of a cell, nor any of cellular functions beyond the operation of a single ribosome. If we are interested in the  computational efficiency of an entire cell we must consider 
other features of the cell as well. In particular, we must consider cellular functions
that maintain information. To do this requires an assessment of the rate of information damage. 
Specifically, we need to assess the rate of loss of proteins due to damage.

Building on previous efforts which describe the general trends in cellular rates and composition\cite{kempes,kempes2}, bacteria provide an ideal case for understanding the trends in computational efficiency across a range of biological scales. A review of analyses of this issue can be found in the Methods section. The key result \cite{kempes2} is that the number of ribosomes required for the cell to be able to divide, $N_{r}$, is bound by the inequality
\begin{equation}
N_{r}\ge\frac{\bar{l}_{p} N_{p} \left(\frac{\phi }{\mu} +1\right)}{\frac{\bar{r}_{r}}{\mu}-\bar{l}_{r} \left(\frac{\eta}{\mu}+1\right)}
\label{ribo-number}
\end{equation}
where both the specific growth rate, $\mu$, and number of proteins, $N_{p}$ have been shown to scale with overall cell volume, $V_{c}$, \cite{delong,kempes,kempes2} (please see the Methods section and \cite{kempes2} for definitions for and values of the constants in Eq. \ref{ribo-number}). The number of proteins is given by
\begin{equation}
N_{p}=P_{0}V_{c}^{\beta_{p}}
\end{equation} 
where, empirically, $\beta_{p}=0.70\pm0.06$ \cite{kempes2}. In addition, the cross-species trends in ribosomes have been shown to follow the lower bound on $N_{r}$ described above \cite{kempes2}, and this is the relationship that we use here for all further calculations. 

\begin{figure}[h!]
\begin{minipage}[b]{.5\linewidth}
\text{(a)\quad\quad\quad\quad\quad\quad\quad\quad\quad\quad\quad\quad\quad\quad\quad\quad\quad\quad\quad\quad\quad\quad\quad\quad}
\text{}
\centering\includegraphics[width=1.\textwidth]{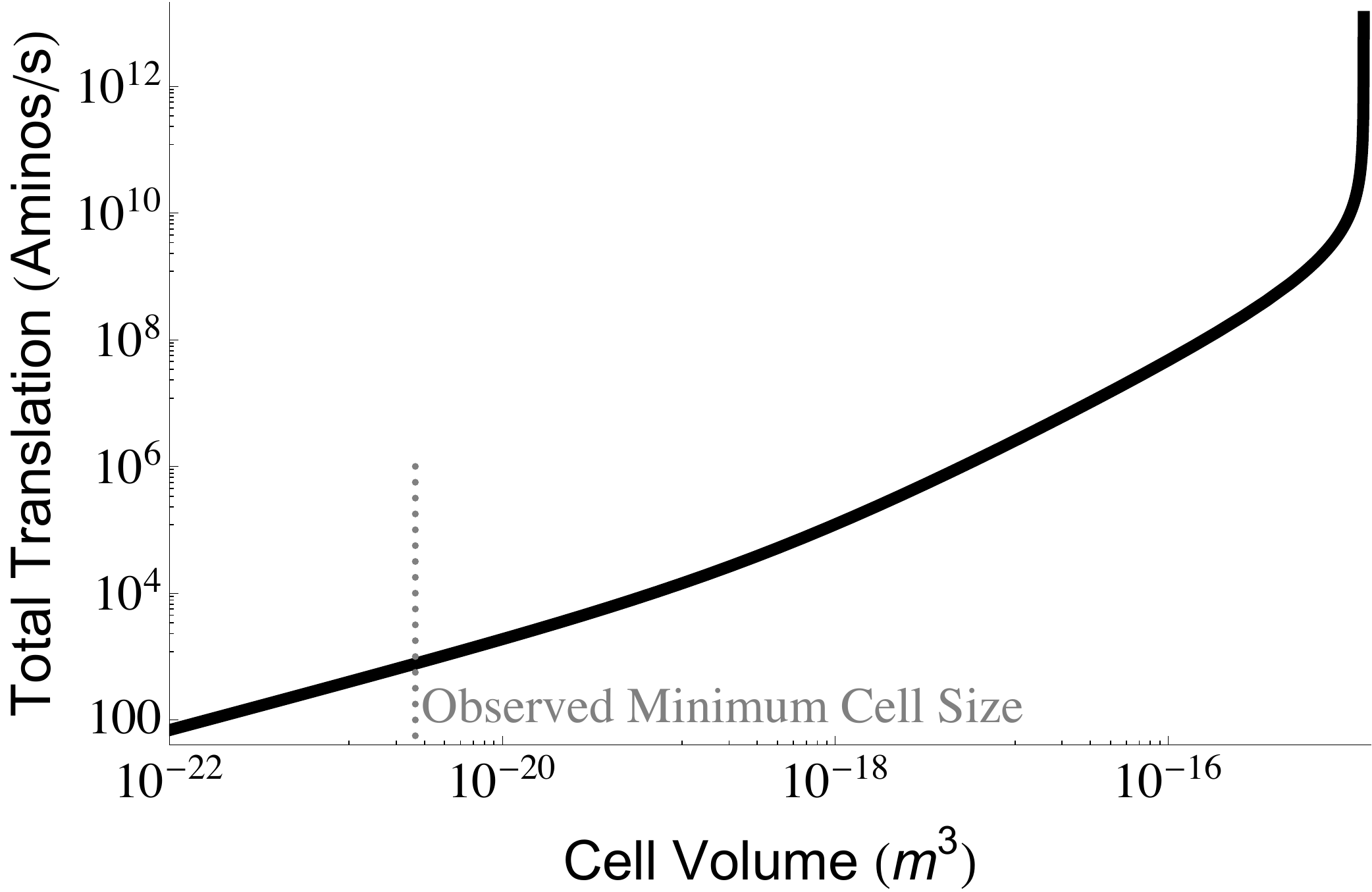}
\end{minipage}
\begin{minipage}[b]{.5\linewidth}
\text{(b)\quad\quad\quad\quad\quad\quad\quad\quad\quad\quad\quad\quad\quad\quad\quad\quad\quad\quad\quad\quad\quad\quad\quad\quad}
\text{}
\centering\includegraphics[width=1.\textwidth]{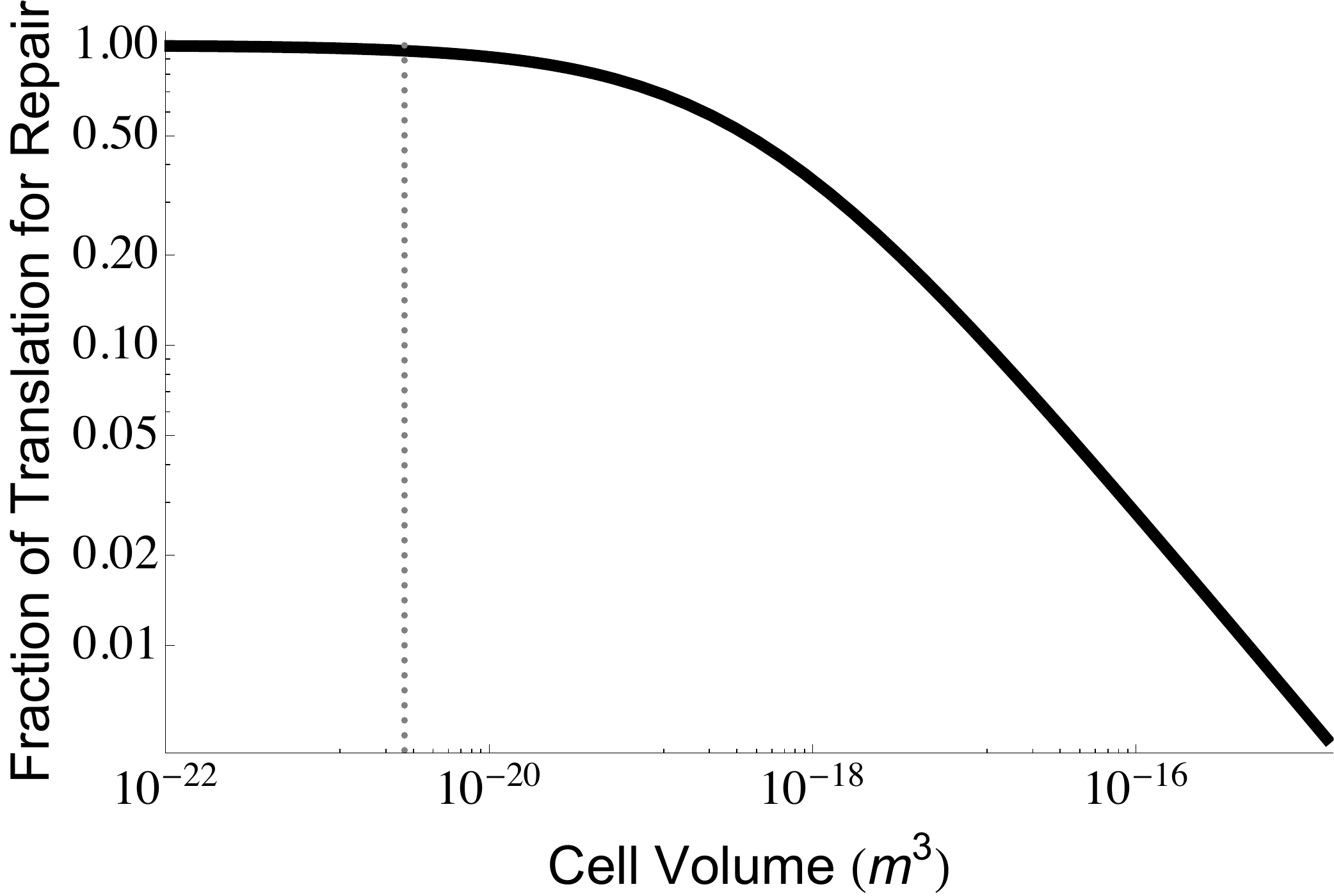}
\end{minipage}
\begin{minipage}[b]{.5\linewidth}
\text{(c)\quad\quad\quad\quad\quad\quad\quad\quad\quad\quad\quad\quad\quad\quad\quad\quad\quad\quad\quad\quad\quad\quad\quad\quad}
\text{}
\centering\includegraphics[width=1.\textwidth]{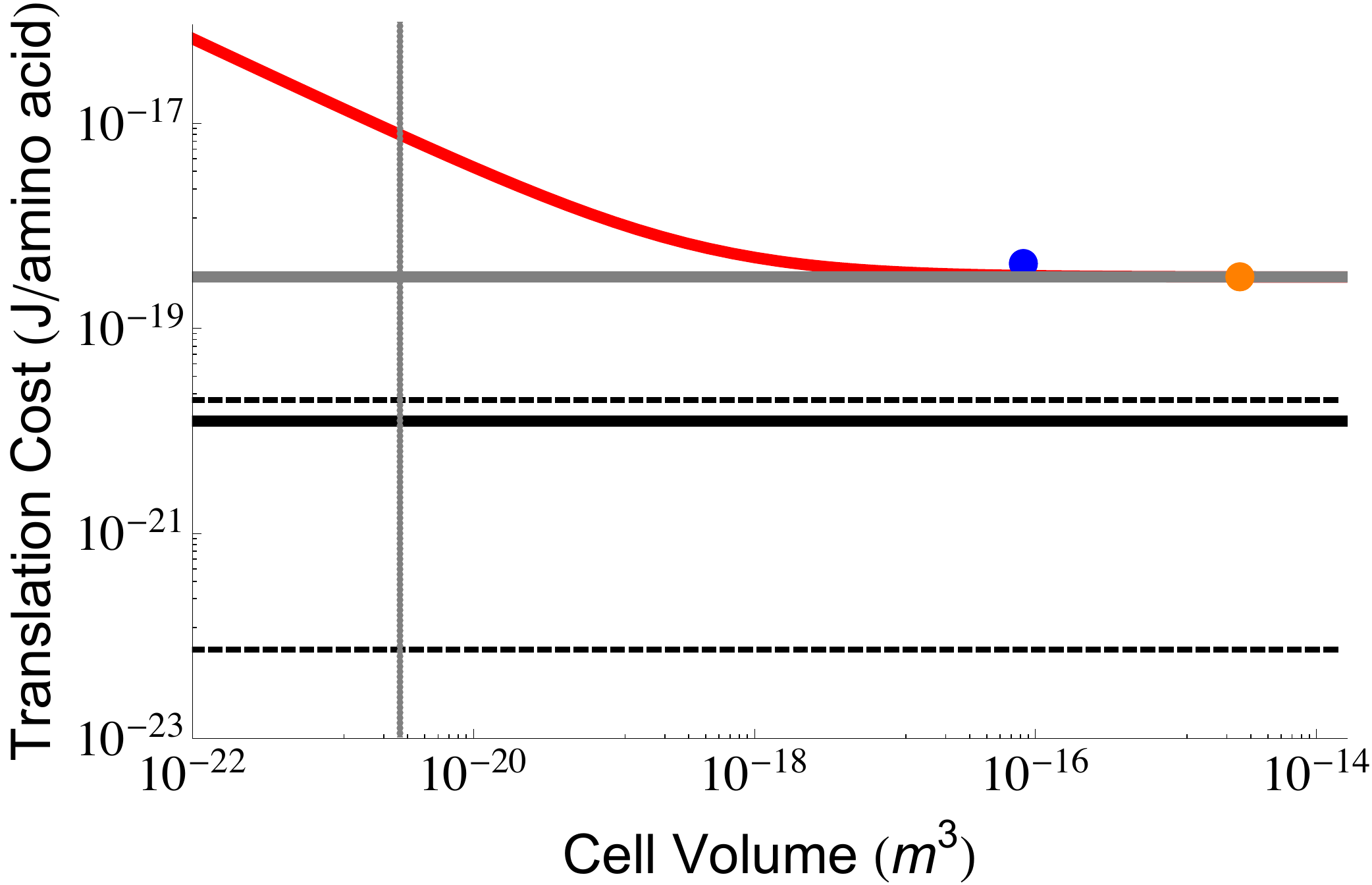}
\end{minipage}
\caption{a.) The total translation rate as a function of bacterial cell volume. The smallest observed bacterial species is indicated with the dotted gray line. b.) The fraction of translation activity that is dedicated to repairing damaged proteins. c.) The useful translation cost, defined as the total energy expended on all translation divided by the number of proteins synthesized for replication rather than for the repair of damage. The red curve is bacteria, the blue point is for unicellular eukaryotes, and the orange point is a single mammalian cell. The black curve is the Landauer bound for translating a protein scaled to a single amino acid addition, and the gray curve is the known energetic cost per amino acid for a single ribosome given the ATP costs. The dashed black lines are the range of Landauer estimates given different accountings of the entropy discussed in the text. For example, the upper dashed line is the case where every amino acid is distinguishable. The discrepancy between the red and gray curve at the small end of bacteria is the result of the high fraction of total translation spent on repair for the smallest cells as shown in (b).}
\label{bacteria-translation}
\end{figure}

This same approach allows us to quantify the total translational computation being performed by the cell (defined here as writing the amino acid pool within a cell to specific protein sequences) as 
\begin{equation}
T_{t}=r_{r}N_{r}/3
\label{eq:T_t}
\end{equation}
(measured in amino acids/s; note the division by 3 to convert from base pairs to amino acids). This is plotted in Figure \ref{bacteria-translation}a. The asymptotic behavior for the largest bacteria is due to the ``ribosome catastrophe'' \cite{kempes2}, the point where the scaling of growth rate (as determined by metabolic rate \cite{kempes}) demands a greater ribosomal capacity than can fit in the cell, or equivalently, the point where cellular division rate becomes faster than the rate at which a ribosome can replicate even the ribosomal proteins. 

We can build off these previous analyses, to analyze the rate of translation that is used
to replace damaged proteins (measured in amino acid units), $R_{t}=\eta N_{r} \bar{l}_{r}/3+\phi N_{p}\bar{l}_{p}/3$. Combining
with Eq.~\eqref{eq:T_t}, the fraction of total translation that is used for such repair is
\begin{equation}
R_{t}/T_{t}=\frac{\eta \bar{l}_{r}}{r_{r}}+\frac{\phi \bar{l}_{p}}{r_{r}}\frac{N_{p}}{N_{r}}.
\label{repair-frac}
\end{equation}
The dependence of this ratio on cell size provides a perspective on the limits of cell size. 
(See Figure \ref{bacteria-translation}b.) At the smallest end of life cellular translation is dominated by the replacement of proteins.
This should be compared to previous results which found that
\emph{total metabolism} is dominated by maintenance processes at the small end of bacteria \cite{kempes}. Here
we tease apart that earlier result, to show that even the individual process of translation 
becomes dominated by maintenance (protein replacement) at the small end of the scale.

As a complement to analyzing the fraction of translation dedicated to repair, we can analyze the 
\textbf{useful} translation, which we define as the accumulated translation that will eventually allow the cell to divide, $U_{t}=\dot{N_{p}}\bar{l}_{p}/3+\dot{N_{r}}\bar{l}_{r}/3$. A convenient way to quantify the overall energetic efficiency of translation is
the ratio of the total energetic cost of translation relative to the useful translation (J per non-degraded amino acid), 
\begin{equation}
E_{t}T_{t}/U_{t}=E_{t}\left(1-\frac{\eta \bar{l}_{r}}{r_{r}}-\frac{\phi \bar{l}_{p}}{r_{r}}\frac{N_{p}}{N_{r}}\right)^{-1}
\label{net-trans-efficiency}
\end{equation}
where $E_{t}$ is the energy to polymerize one amino acid ($4$ ATP at $47.7$ (kJ/mol) of ATP). In Figure \ref{bacteria-translation}c we have plotted this overall efficiency of translation based on the known scaling of $N_{r}$ and $N_{p}$ across bacteria. We find that the smallest cells are over an order of magnitude less energetically efficient than the largest cells at performing the operation of adding an amino acid to the protein chain. The largest cells converge to the energetic efficiency of the ribosome itself, which is still about 26 times larger than the Landauer bound as discussed earlier (Figure \ref{bacteria-translation}c). 

Another interesting question is how the translational efficiency changes across diverse biological architectures. We find that the translational efficiency of single-cell eukaryotes (considering values for yeast; see methods) and single mammalian cells does not significantly deviate from the efficiency of a single ribosome (Figure \ref{bacteria-translation}c) which is also the efficiency of the largest bacteria. Life quickly converges to the efficiency of the ribosome as cells become larger, and maintains that efficiency at the cellular level across the diversity of both free-living and multicellular eukaryotes. It should be noted that a critical feature of both Equation \ref{repair-frac} and Equation \ref{net-trans-efficiency} is the ratio of $N_{p}$ to $N_{r}$. In particular, the global cost of useful translation and the fraction of translation dedicated to repair are minimized by $N_{p}/N_{r}=0$.  Our results show that after a sufficient cell size, life has been able to adjust this ratio such that the effective translational cost is only negligibly larger than that of the ribosome. A surprising result here is that while the bulk power consumption of organisms dramatically shifts across major evolutionary transitions, the unit costs of translation are held constant once cells reach a sufficiently large size. 

\subsection{Rates of cellular computation measured in Oklos}

So far we have only considered one aspect of the thermodynamic cost of cellular computation,
namely energy spent per amino acid operation, measured either
using the Landauer bound or the efficiency achieved by a single ribosome. However there are other important metrics 
for analyzing cellular computation.
One that was recently introduced is the \textbf{Oklo}, defined as number of bit operations per gram per second \cite{laughlin}. 

\begin{figure}[h!]
\begin{minipage}[b]{.5\linewidth}
\text{(a)\quad\quad\quad\quad\quad\quad\quad\quad\quad\quad\quad\quad\quad\quad\quad\quad\quad\quad\quad\quad\quad\quad\quad\quad}
\text{}
\centering\includegraphics[width=1.\textwidth]{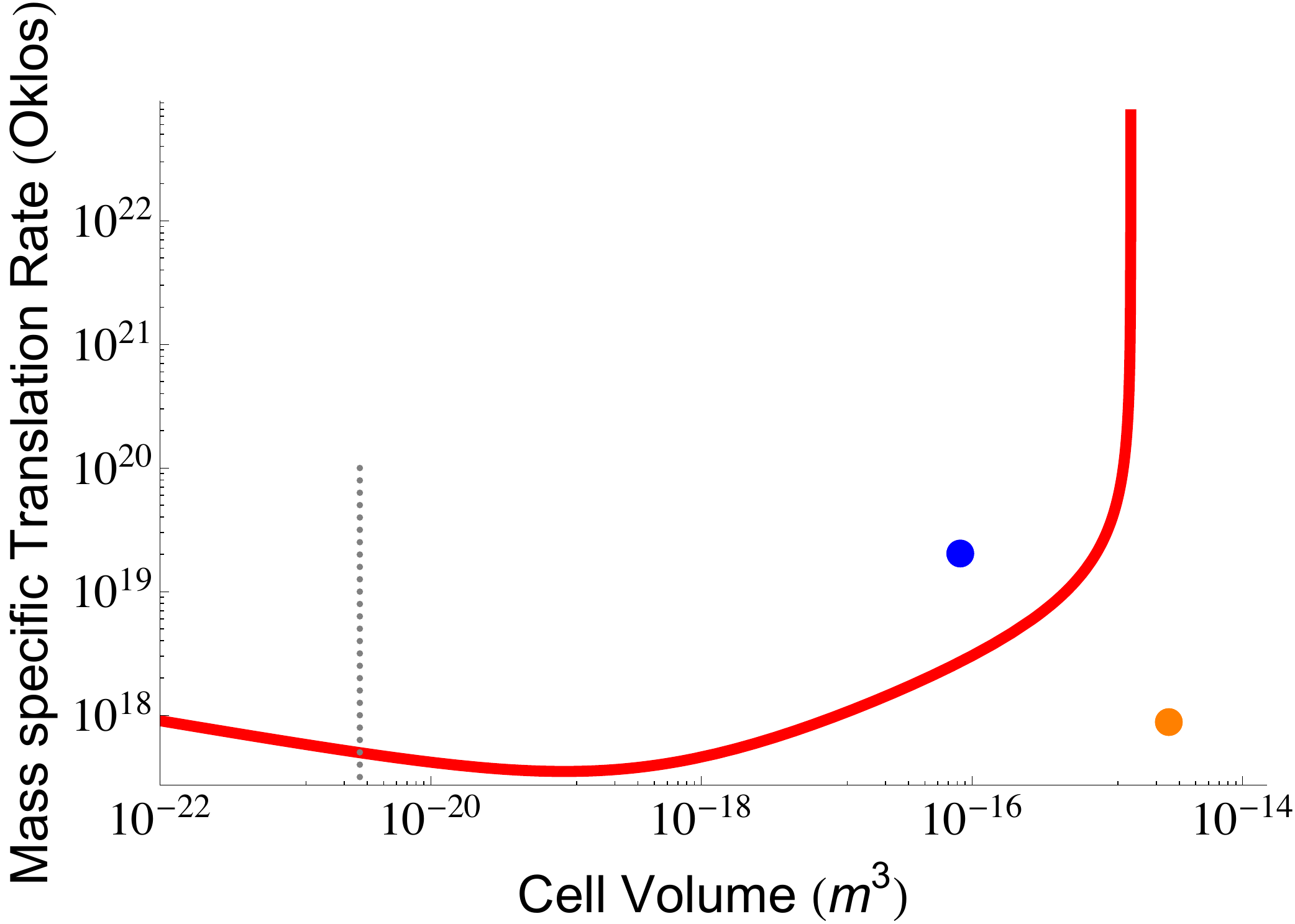}
\end{minipage}
\begin{minipage}[b]{.5\linewidth}
\text{(b)\quad\quad\quad\quad\quad\quad\quad\quad\quad\quad\quad\quad\quad\quad\quad\quad\quad\quad\quad\quad\quad\quad\quad\quad}
\text{}
\centering\includegraphics[width=1.\textwidth]{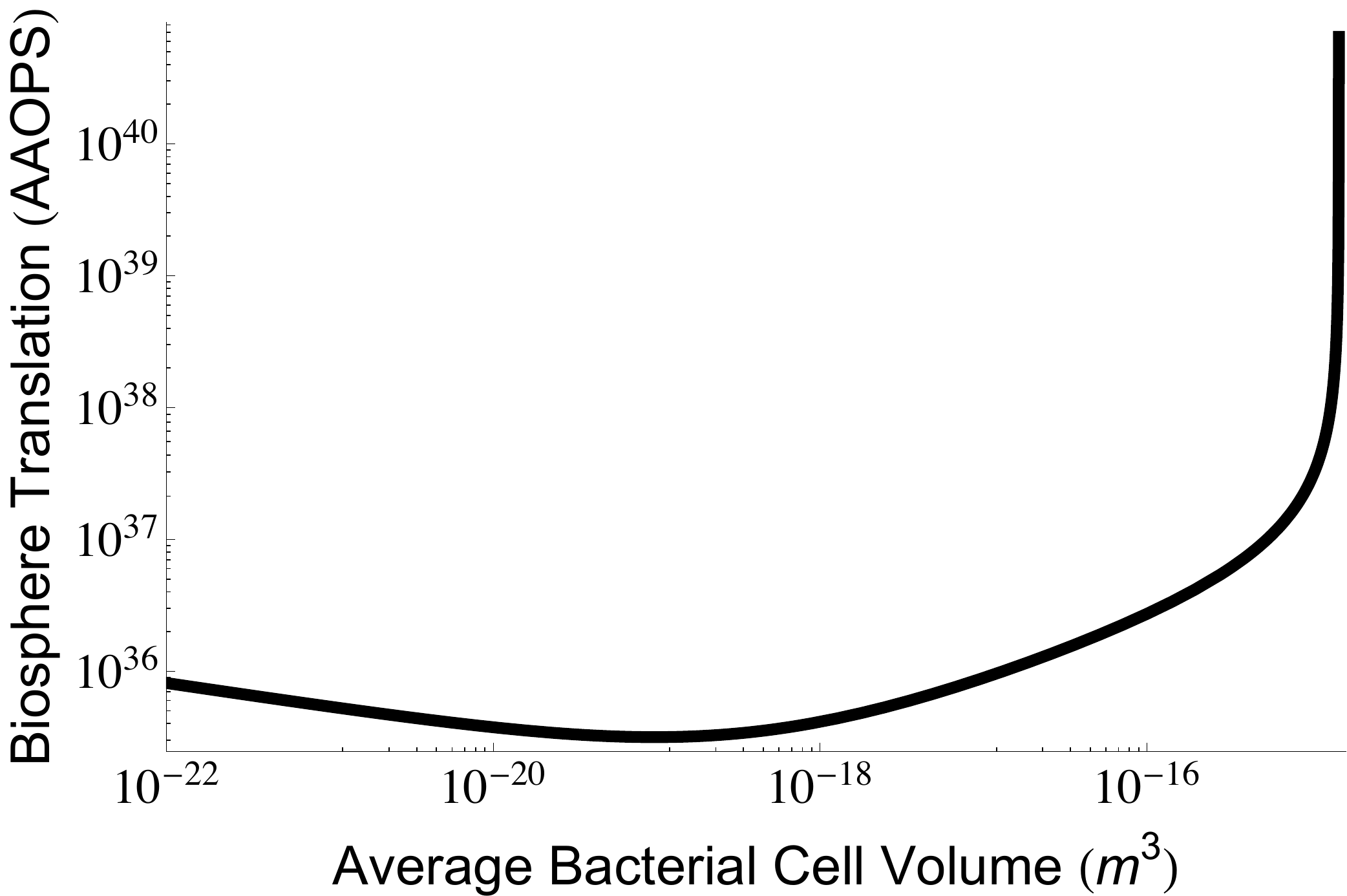}
\end{minipage}
\caption{a.) The mass specific translation rate of individual cells of different size in Oklos (here considering amino acids per gram per second). The red curve is the average cross-species relationship calculated for bacteria, the blue point is for yeast, and the orange point is a mammalian cell. b.) The total amino acid operations per second (AAOPS) of all the bacteria in the biosphere as a function of the average bacterial cell volume. The dashed gray line indicates the smallest observed bacterial species.}
\label{Oklos-figure}
\end{figure}

Our analysis above allows us to calculate one component of the number of oklos expended in cellular computation: the rate of amino acid bit operations per unit of bacterial mass. 
We plot this as a function of the overall size of the bacterial species in Figure \ref{Oklos-figure}a. We find a non-monotonic function for bacteria with a minimum in the mid range of bacterial sizes, and a rapid increase for the largest bacteria due to the increased number of ribosomes \cite{kempes2}. However, it should be noted that for most bacterial sizes the Oklos curve is surprisingly flat and ranges within an order of magnitude. The unicellular eukaryote value from yeast is significantly larger than that of bacteria of the same size, and is about an order of magnitude larger than the value for bacteria of average cell size. However, the mammalian cell values appear to be indistinguishable from the value for a bacterium of average size. In contrast to the energetic efficiency of translation, which appears to saturate at the ribosome minimum and be held constant across evolutionary transitions, the Oklos found within different biological architectures have significant shifts across diverse life, where unicellular eukaryotes are able to achieve the highest rate of bit operations per unit mass. 

\subsection{Translational computation by the biosphere}

Known cellular rates of the amount of computation per unit mass have recently been scaled up 
to analyze the rate of translational computation performed by the biosphere \cite{laughlin}. Having done this, one can then
divide by total solar flux to calculate the
thermodynamic efficiency of translation at the scale of the biosphere \cite{laughlin}. However, given the strong scaling relationships demonstrated 
both in this paper and in previous work \cite{delong,kempes,kempes2}, it is important to note that any such estimates depend strongly on how we model cell size distributions in different environments. In addition, we have seen that growth rate plays an important role in determining the overall computational efficiency of translation, and it should be noted that large amounts of the biomass on earth is growing at rates close to zero \cite{hoehler}. These low growth rates will not affect the assessment of the information stored in the biosphere in DNA \cite{landenmark}, but will matter for assessing the overall rates of computation. 

To begin to address these subtleties we have plotted (Figure \ref{Oklos-figure}b) the total translation of bacteria (at maximum growth rate; measured in amino acid operations per second ``AAOPS'') based on the previous estimate of total bacterial cells in the biosphere \cite{landenmark}. These results show that the total computation rate of translation in the biosphere could range over many order of magnitude depending on average cell size and growth rate (which scales with cell volume in these calculations). Nonetheless, the 
estimate of the amount of computation occurring in translation (measured in AAOPS) greatly exceeds previous estimates of the nucleotide operations per second, which range between $10^{24}$ and $10^{29}$ \cite{landenmark,laughlin}. However, it should be noted that this estimate is based on the scaling of cells growing at maximum rate which does not represent most of the Earth's biomass \cite{hoehler}. Following \cite{laughlin}, and considering the slowest growing bacteria \cite{hoehler} we can calculate an extreme lower bound on the computational efficiency of the biosphere by assuming that all usable sunlight is dedicated to the total translation. We find that the biosphere would have a total of $1.86\times10^{30}$ (AAOPS/biosphere) leading to an efficiency of $3.65\times10^{13}$ (AAOPS/J of sunlight). This value is close to but exceeds the estimate of $2\times10^{12}$ (bit operations/J of sunlight) from \cite{laughlin} for DNA replication. 

\subsection{DNA replication efficiency}

\begin{figure}[h!]
\begin{minipage}[b]{.5\linewidth}
\text{(a)\quad\quad\quad\quad\quad\quad\quad\quad\quad\quad\quad\quad\quad\quad\quad\quad\quad\quad\quad\quad\quad\quad\quad\quad}
\text{}
\centering\includegraphics[width=1.\textwidth]{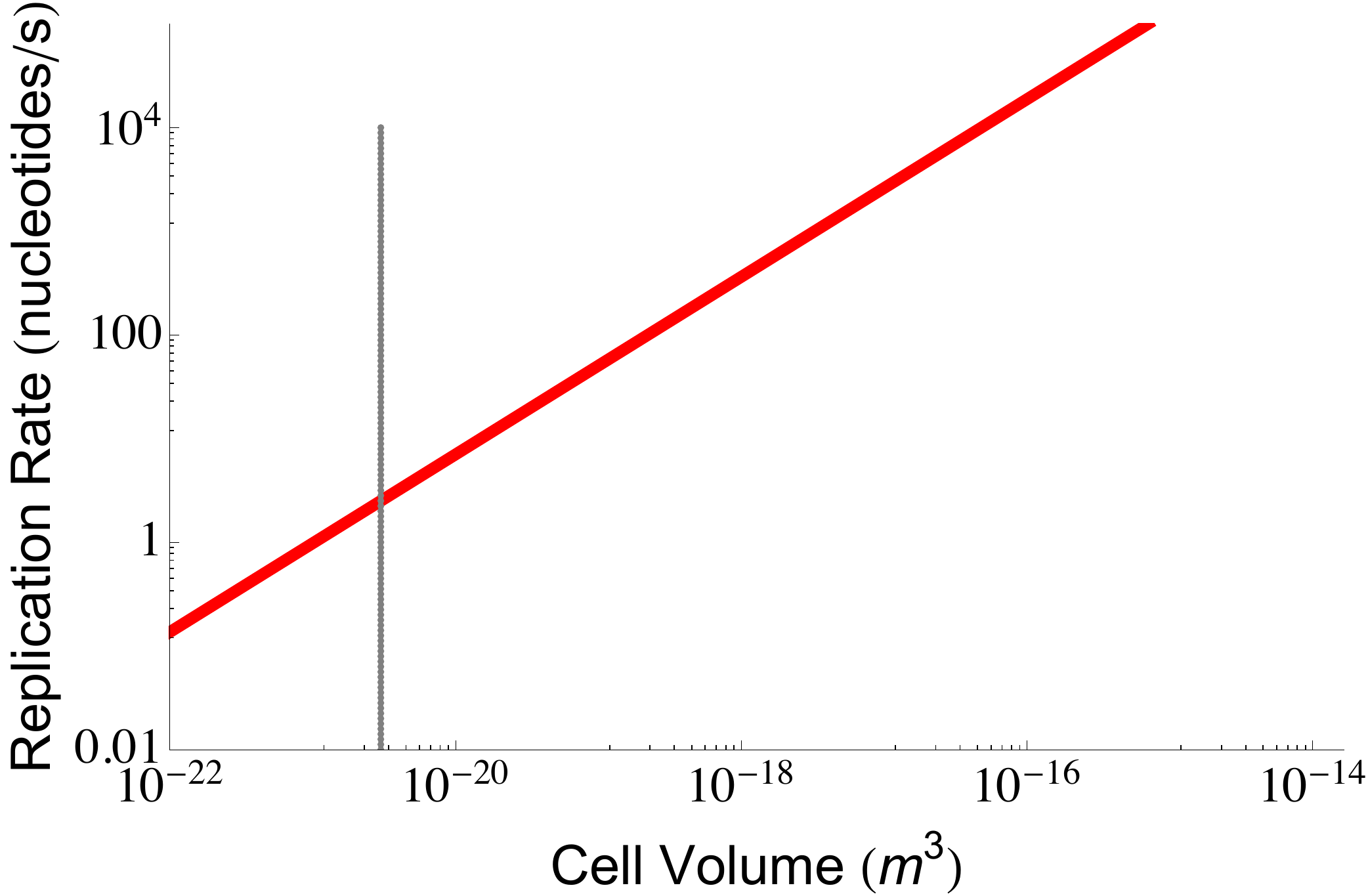}
\end{minipage}
\begin{minipage}[b]{.5\linewidth}
\text{(b)\quad\quad\quad\quad\quad\quad\quad\quad\quad\quad\quad\quad\quad\quad\quad\quad\quad\quad\quad\quad\quad\quad\quad\quad}
\text{}
\centering\includegraphics[width=1.\textwidth]{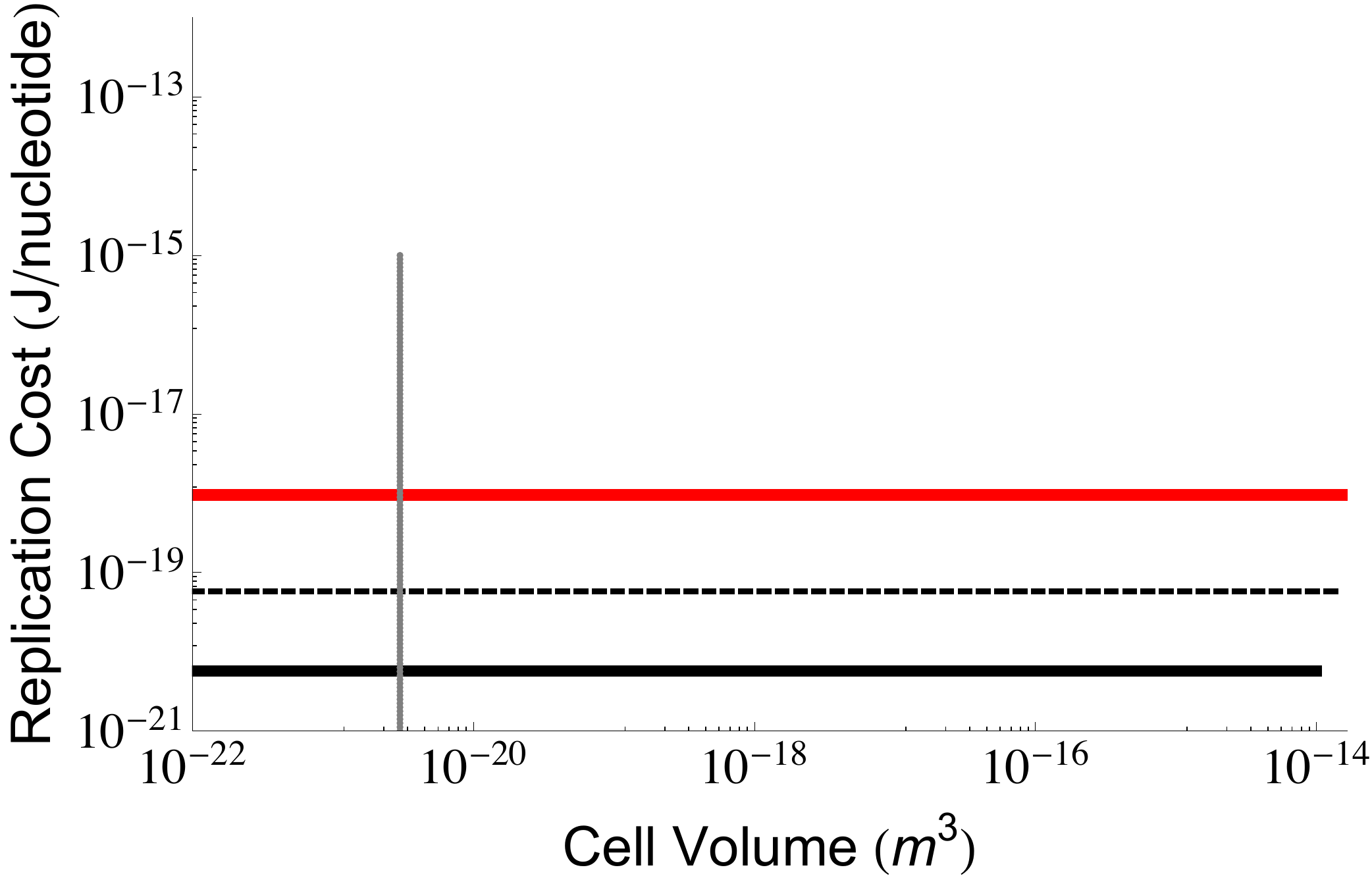}
\end{minipage}
\begin{minipage}[b]{.5\linewidth}
\text{(c)\quad\quad\quad\quad\quad\quad\quad\quad\quad\quad\quad\quad\quad\quad\quad\quad\quad\quad\quad\quad\quad\quad\quad\quad}
\text{}
\centering\includegraphics[width=1.\textwidth]{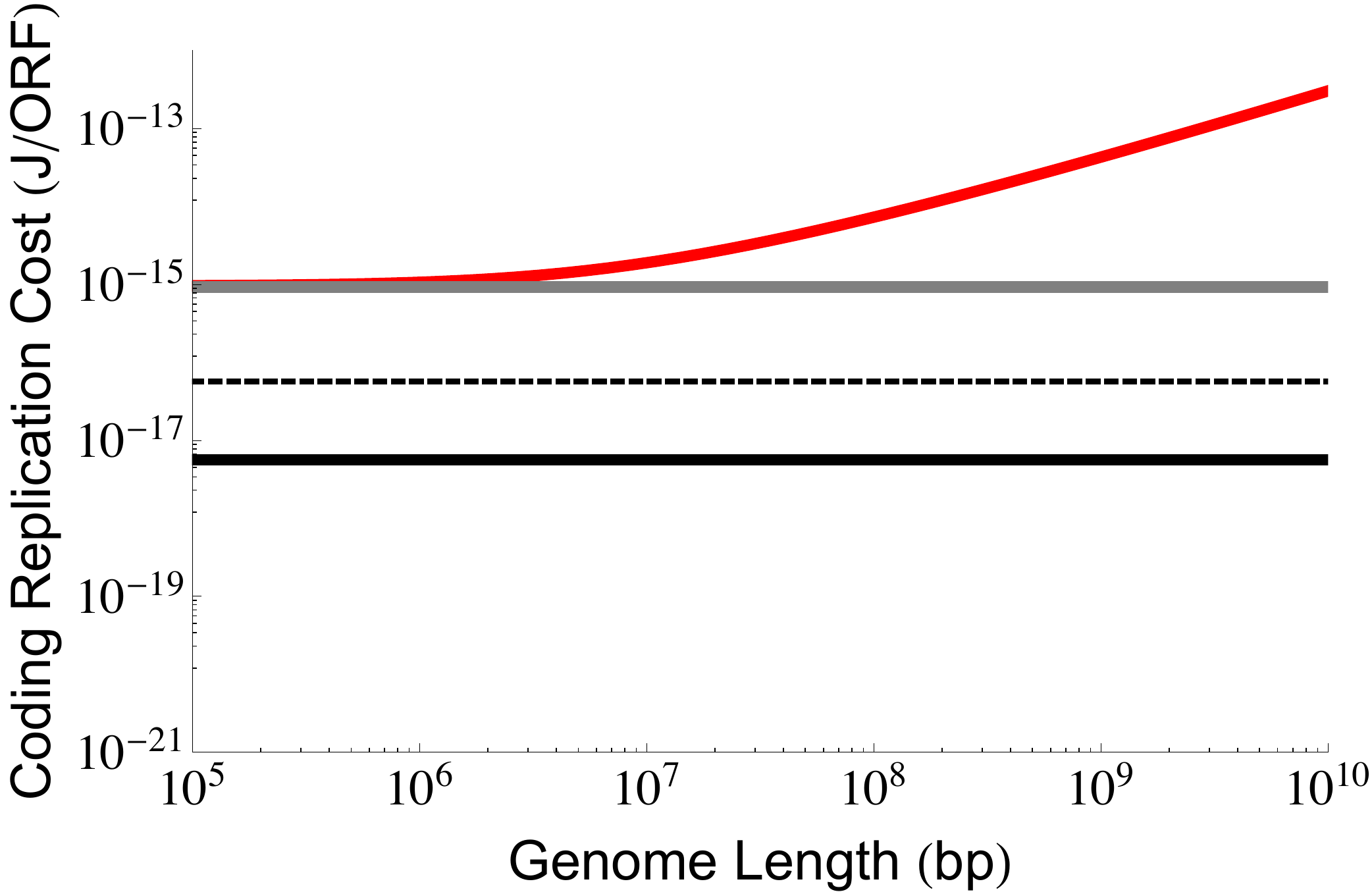}
\end{minipage}
\caption{ a.) The total DNA replication rate as a function of bacterial cell volume. The smallest observed bacterial species is indicated with the gray dotted line. b.) The DNA replication efficiency compared with the Landauer bound for copying a single nucleotide. c.) The DNA replication efficiency for open reading frames considering the total cost of replication. The Landauer bound for copying a single nucleotide has been scaled up using the average length of a gene.}
\label{bacteria-DNA-replication}
\end{figure}

As noted above, much of the previous work on biological computation has focused on the process of DNA replication \cite{landenmark,laughlin}. Accordingly, in parallel to our analysis of translation, we consider how the total computations and efficiencies of DNA replication shift across bacteria of different size. In Figure \ref{bacteria-DNA-replication}a we have plotted the total nucleotide replication rate for bacterial species of different size (to form this figure we multiply overall genome size by division rate, each of which follows a known scaling relationship with cell volume \cite{kempes,kempes2}). We find that the rate of nucleotide copying varies by several orders of magnitude across the range of bacteria.

Paralleling our calculations of string writing during translation we can use similar entropic considerations to estimate the Landauer bound in DNA replication. For a uniform nucleotide bath, the number of states is $m=4^{G}$, where $G$ is the length of a genome, and the corresponding Landauer bound is $1.86\times10^{-14}$ (J) for a typical bacterial genome size, which can more meaningfully be converted to $5.74\times10^{-21}$ (J/nucleotide), a value that will not vary with genome size. The known value of $12$ ATP per nucleotide copying in cells gives a value of $9.50\times10^{-19}$ (J/nucleotide) which is $165$ times larger than the Landauer bound. Thus, bacteria consume about two orders of magnitude more energy than the Landauer bound for DNA replication. 

However, just as we saw with translation, there is a broad range of possibilities for the details of the string writing that affect the associated calculations of the entropies. This is particularly relevant for DNA replication, where it is much more likely that the string is written from a pool of nucleotides that approximately match the length and composition of the genome. As described earlier, this scenario would give $m=G!/\left[\left(G/4\right)!\right]^{4}$, leading to a Landauer bound of $5.74\times10^{-21}$ (J/nucleotide), which is indistinguishable from the value above. For a set of completely distinguishable nucleotides we would have $m=G!$ giving $5.80\times10^{-20}$ (J/nucleotide) which is an order or magnitude larger than the other thermodynamic cost estimates for DNA replication. This range of Landauer bound estimates is shown in Figures \ref{bacteria-DNA-replication}b-c.

\subsection{Thermodynamic efficiency of gene replication}
While the unit costs of replicating a nucleotide are not changing across cells of different size, it should be noted that across unicellular bacteria and eukaryotes the percentage of the genome dedicated to coding regions is decreasing with increasing genome size \cite{ahnert,friar}. Specifically, it has been shown that a good empirical fit to the number of open reading frames (ORFs) is given by 
\begin{equation}
ORF\left(G\right)=A\ln\left(1+\frac{G}{B}\right)
\label{benfordorf}
\end{equation}
where $A = 4016\pm280$ ORF, $B=4106\pm680$ kbp and $G=cDNA+ncDNA$, with $cDNA$ and $ncDNA$ denoting the coding and non-coding fractions of the genome \cite{friar}. This relationship makes it possible to quantify the cost of replicating a coding nucleotide relative to the Landauer bound. Figure \ref{bacteria-DNA-replication}c shows that the total cost for replicating a gene is increasing as genomes become larger. This result is the opposite of what we found in protein translation where the smallest and simplest cells are the least efficient at translation at a whole cell-level because of the high overhead of protein replacement. 

There has been recent interest in understanding how the energetics of single genes change across the range of life \cite{lane,lynch,kempes2,kempes3}. For example, it has been shown that the fraction of the total energy budget spent on DNA replication is decreasing with increasing cell size \cite{lynch}. However, the results here show that compared with an absolute unit efficiency, the replication cost for a single nucleotide from a coding region is increasing. It is possible that the decreasing relative cost of replication compared to total metabolism as cell size increases allows for this inefficiency. It has also been proposed that the underlying distribution leading to Equation \ref{benfordorf} is a Benford distribution, and that this gives genomes the following properties: (a) upon combination of genes the minimal error possible is made (maximum fidelity) and (b) the information contained in the genes is transferred at the maximum possible rate (minimizing distortion) \cite{friar,friar2,ciofalo}. Thus far in this paper we have analyzed the thermodynamic efficiency of the computational processes, however, the above connection opens up important future efforts which should focus on the connection between the thermodynamic efficiency of both computation and communication within cells.

\subsection{DNA computation and storage of the biosphere}

In addition to the overall computational rate in DNA replication, another important characteristic of 
naturally occurring DNA computation is the total storage capacity of DNA,
both within a single cell and within the biosphere as a whole. We find that these values again depend strongly on assumptions about average bacterial size. In Figures \ref{bacteria-DNA-replication}c and \ref{bacteria-DNA-replication}d we have plotted both the storage of a single cell and of the total bacterial biomass in the biosphere as a function of cell size. Each varies over about an order of magnitude. We note that the calculation for the biosphere would agree with the previous estimate of $1.6\times10^{37}$ (bp) \cite{landenmark} only for an assumption that most of the biomass in the biosphere is small bacteria. Similar to our calculation of the amino acid operations per total energy of the biosphere, above we calculate that there are $2.54\times10^{29}$ (nucleotide operations/s/biosphere) or $4.98\times10^{12}$ (nucleotide operations/J) which agrees well with the independent estimate from \cite{laughlin}.

\begin{figure}[h!]
\begin{minipage}[b]{.5\linewidth}
\text{(a)\quad\quad\quad\quad\quad\quad\quad\quad\quad\quad\quad\quad\quad\quad\quad\quad\quad\quad\quad\quad\quad\quad\quad\quad}
\text{}
\centering\includegraphics[width=1.\textwidth]{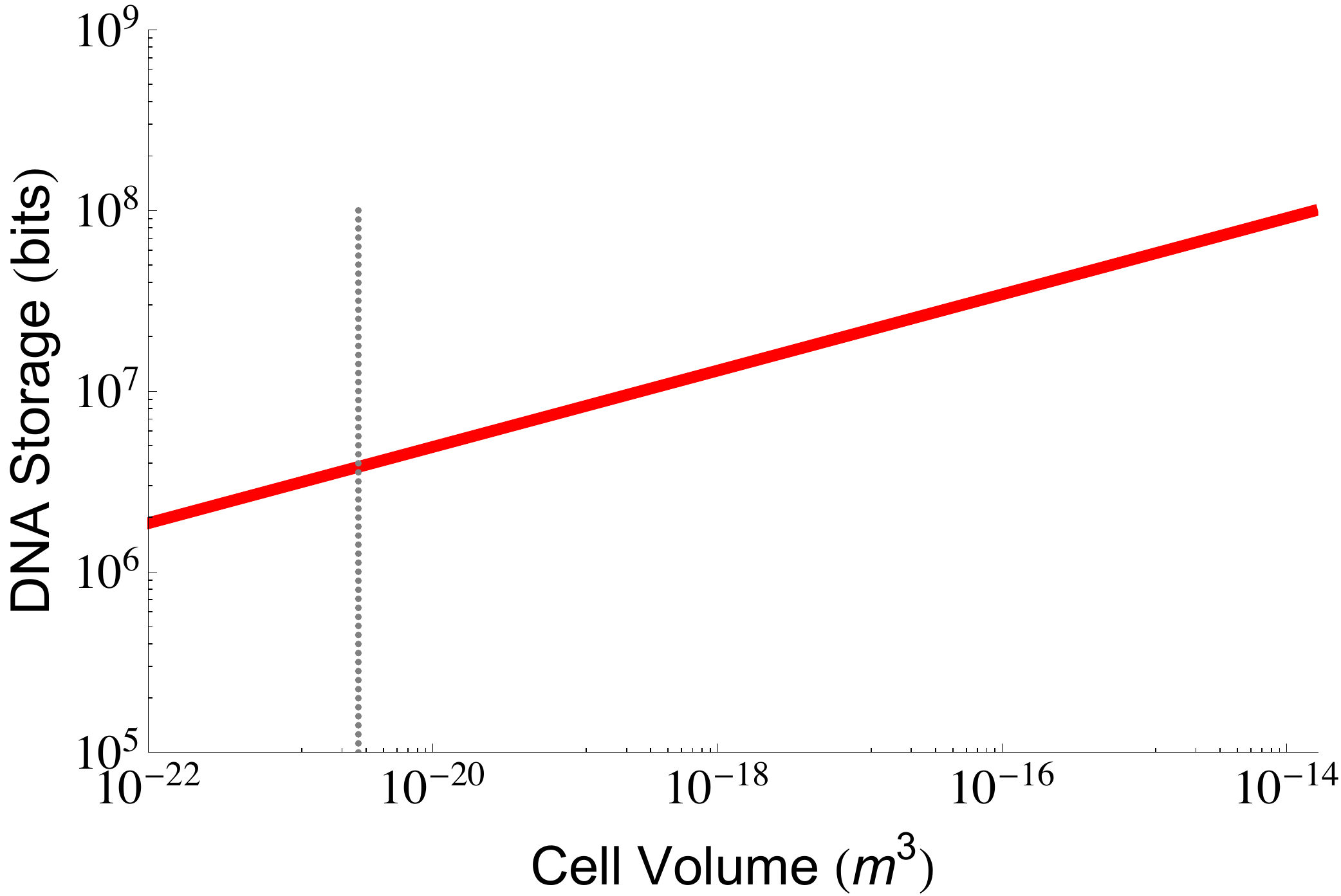}
\end{minipage}
\begin{minipage}[b]{.5\linewidth}
\text{(b)\quad\quad\quad\quad\quad\quad\quad\quad\quad\quad\quad\quad\quad\quad\quad\quad\quad\quad\quad\quad\quad\quad\quad\quad}
\text{}
\centering\includegraphics[width=1.\textwidth]{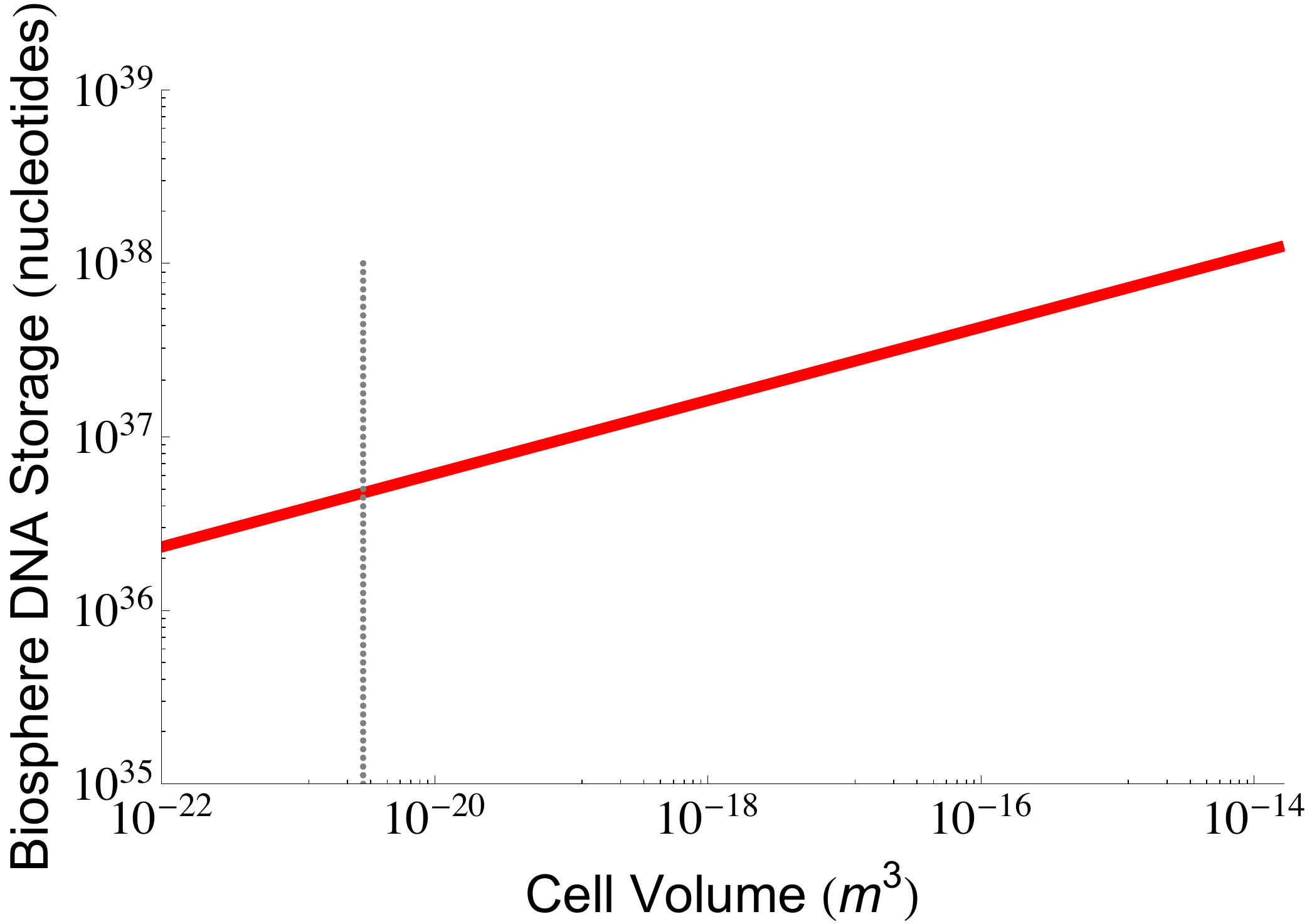}
\end{minipage}
\caption{ a.) The storage capacity of the DNA in a single cell as a function of total cell size and b.) these same values scaled up to the biosphere.}
\label{bacteria-DNA-storage}
\end{figure}

\section{Discussion}

Here we have shown that life maintains a roughly constant power usage as a function of the age of the system's evolutionary arrival, yet the overall scale and type of system has strong implications for power usage: bacteria increase in power usage per unit mass for larger cells, whereas multicellular life has a decreasing power expenditure per unit mass for larger systems. In fact, multicellular life would be surprisingly more efficient than astronomical objects if extrapolated to the same scales. 

Despite these shifts across the architectures of life, we find surprising consistency in the efficiency of translation, one of
the most universal types of computation carried out in biological systems. Our analyses show that as bacteria become larger their 
overall translational efficiency converges on that of a single ribosome. In
addition, this efficiency is maintained for unicellular eukaryote and mammalian cells. Astonishingly, this efficiency is only about an order of magnitude larger than the Landauer bound, and is an impressive feat of biology as it far exceeds modern computers. However to properly
``calibrate'' this efficiency we would need to know how close to the Landauer bound biology could have gotten
using alternate biochemical processes (arrived at via alternative evolutionary histories) to perform translation. On the
other hand, it is important to note that the processes considered here represent only a fraction of the total computations of the cell.
 In the future it will be important to quantify the computational efficiency of various levels of biological physiology. These additions should range from metabolic networks in bacteria, to chromatin computations in unicellular eukaryotes (e.g. \cite{prohaska}), to the social computations of multicellular mammals \cite{flack, flack2}, where each new level of hierarchy integrates the computations of the lower levels \cite{flack}. In addition, we note that our calculations underestimate the full computational cost of translation, since they treat amino
acids as one-dimensional strings, ignoring the computational cost of reducing the three dimensional positional entropy of amino acids into
a single string. Accordingly, they underestimate the thermodynamic efficiency of biological computation which is already impressively close to the lower bounds considered here. It should also be noted that ATP are used for many cellular processes in addition to those considered here. Understanding the thermodynamic efficiency for these other enzymatic and metabolic processes will require quantifying the computations being performed, and this represents a major challenge of future interest for the community.

Furthermore, we have shown how the overall computational efficiency of translation in the biosphere greatly depends on how much of the total biomass is partitioned into organisms of different size. This type of analysis of the biosphere, similar to previous efforts \cite{landenmark,laughlin}, provides new ways to quantify ecological efficiency. Yet it should be noted that a huge fraction of biomass on earth is partitioned into the smallest cells which are the least computationally efficient from the perspective of translation. This implies that despite the
evident efficiency of the biosphere, it could have been higher, and is clearly not the dominant force for evolutionary selection in some systems. On the other hand, it may well be that it is not possible to maintain an entire
biosphere \emph{without} maintaining a level of diversity that results in lots of organisms with inefficient computation.
Under that hypothesis, it may well be that the biosphere as a whole computes with close to the maximal possible
thermodynamic efficiency. 

\section{Methods}

\subsection{Review of ribosome requirements}
Previous analyses have shown that the number of ribosomes can be predicted based on the overall growth rate of cells and the total protein content \cite{kempes2}. These analyses were based on the translation and degradation dynamics given by
\begin{equation}
\dot{N_{r}}=\gamma\frac{r_{r}N_{r}}{\bar{l}_{r}}-\eta N_{r}
\label{}
\end{equation}   
\begin{equation}
\dot{N_{p}}=\left(1-\gamma\right)\frac{r_{r}N_{r}}{\bar{l}_{p}}-\phi N_{p}
\end{equation}   
where it has been shown \cite{kempes2} that the partitioning of translation between ribosomal and non-ribosomal proteins is bounded by
\begin{equation}
\gamma\ge\frac{\bar{l}_{r} \left(\eta  t_{d}+\ln (2)\right)}{\bar{r}_{r} t_{d}} 
\end{equation}
where $t_{d}=\ln (2)/\mu$ is the division time, $\bar{l}_{r}$ is the total length of all ribosomal proteins in base pairs as a cross-species average, $\bar{l}_{p}$ (bp) is the average length of all other cellular proteins,  $\bar{r_{r}}=63$ (amino acids s$^{-1}$) \cite{bremer} is the maximum base pair processing rate of the ribosome, $\eta$ (s$^{-1}$) and $\phi$ (s$^{-1}$) are specific degradation rates for ribosomes and proteins respectively (both taken to be $6.20\times10^{-5}$ \cite{kempes2}), $\mu$ is the specific growth rate, and $N_{p}$ is the total number of proteins. 

\begin{table}[!h]
\caption{Definitions of parameters and constants}
\label{param_table}
\begin{tabular}{lll}
\hline
Parameter & Units & Definition  \\ 
\hline
$t_{d}$ & (s) & Division Time \\
$\bar{l}_{r}$ & (bp) & Length of all ribosomal proteins \\
$\bar{l}_{p}$ & (bp) & The average length a protein \\
$\bar{r_{r}}$ & (bp $\cdot$ s$^{-1}$) & Ribosomal base pair processing rate \\
$\eta$ & (s$^{-1}$) & Specific ribosome degradation rate \\
$\phi$ & (s$^{-1}$) & Specific ribosome degradation rate \\
$\mu$ & (s$^{-1}$) & Specific growth rate \\
$N_{p}$ & & Total number of proteins \\
$N_{r}$ & & Total number of ribosomes \\
$\gamma$ & & Fraction of translation dedicated to making ribosomes  \\
$V_{c}$ & (m$^{3}$) &Cell volume \\
$R_{t}$ & (amino acids $\cdot$ s$^{-1}$) & Repair translation \\
$T_{t}$ & (amino acids $\cdot$ s$^{-1}$) & Total translation \\
$U_{t}$ & (amino acids $\cdot$ s$^{-1}$) & Useful translation \\
$E_{t}$ & (J $\cdot$ amino acids$^{-1}$) & Energy per amino acid polymerization \\
\hline
\end{tabular}
\end{table}

\subsection{Parameter values}

In Figure \ref{power-density} the lines represent transformations (division by cell volume) of the best OLS best fits of metabolic rate against cell volume (compared with the RMA fits carried out in \cite{delong}). The large scatter in the data for unicellular eukaryotes is due to the fact that the metabolic rate scaling is approximately linear with cell volume, and thus power density is effectively just the residual values around this linear scaling relationship. 

For the translational efficiency in unicellular eukaryotes we combined values for yeast and for multicellular eukaryotes we used values for mammalian cells. The equations for bacteria are general provided that we can accurately estimate the number of ribosomes and proteins, and the overall time to divide for a given cell volume. For yeast we use a value of $N_{r}=1.87\times10^{5}$ \cite{harr}, $N_{p}=5\times10^{7}$ \cite{futcher}, and a division time of $t_{d}=7561.6$ (s) \cite{johnston}. For mammalian cells we used $N_{r}=1.27\times10^{7}$ \cite{weibel}, $N_{p}=1.70\times10^{5}$ \cite{li}, and a division time of $t_{d}=1.71\times10^{5}$ (s) \cite{baserga}.

\section*{Acknowledgment}

C.P.K. acknowledges the support of the Omidyar Fellowship at the Santa Fe Institute.
D.H.W. would like to thank the Santa Fe Institute for helping to support this research. This paper was made possible through the support of Grant No. TWCF0079/AB47 from the Templeton World Charity Foundation, Grant No.
FQXi-RHl3-1349 from the FQXi foundation, and Grant No. CHE-1648973 from the U.S. National Science Foundation. 
The opinions expressed in this paper are those of the authors and do not necessarily 
reflect the view of Templeton World Charity Foundation.

\end{document}